\documentstyle[aps,graphicx,epsfig]{revtex}\tighten
\draft

\def\be{\begin{equation}}
\def\ee{\end{equation}}
\def\bea{\begin{eqnarray}}
\def\eea{\end{eqnarray}}

\def\gam{\gamma}

\def\Om{\Omega}

\def\e{{\rm e}}

\def\ti{\tilde}
\def\hsp5{\hspace{5mm}}

\def\case#1/#2{\textstyle\frac{#1}{#2}}

\begin{document}

\title{Isotropic singularity in  brane cosmological models}

\author{Alan Coley}
\address{Department of Mathematics \& Statistics,
Dalhousie University\\
Halifax, Nova Scotia, Canada  \enskip  B3H 3J5}

\address{E-mail: aac@mathstat.dal.ca}

\maketitle

\today
\begin{abstract}
It is argued that the initial cosmological singularity
is isotropic in spatially
inhomogeneous brane-world models. This implies that brane
cosmology may naturally give rise to a set of initial data that
provide the conditions for inflation to subsequently take place,
consequently offering a plausible solution
to the initial conditions problem in cosmology.

\end{abstract}

\pacs{ 98.80.Cq} \vskip2pc

Cosmology models in which our Universe is a four-dimensional brane
embedded in a higher-dimensional spacetime is currently of great
interest. In the Brane-world scenario, ordinary matter fields are
confined to the brane, while the gravitational field can also
propagate in the extra dimensions (i.e., in the
`bulk')~\cite{rubakov}. In particular, generalized
Randall-Sundrum-type models \cite{randall} are relatively simple
phenomenological models which are inspired by string theory. These
models are self-consistent and simple and allow for an
investigation of the essential non-linear gravitational dynamics,
particularly in the high-energy regime close to the initial
singularity, but captures some of the essential features of the
dimensional reduction of eleven-dimensional supergravity
introduced by Ho$\check{\mbox{r}}$ava and Witten~\cite{Horava}.
Randall-Sundrum type models have in common a five dimensional
space-time (bulk), governed by the Einstein equations, and a four
dimensional brane, representing our physical world, on which
ordinary matter fields are confined. At low energies gravity is
also localized at the brane (even when the extra dimensions is not
small) \cite{randall}. There are many generalizations of the
original Randall-Sundrum scenario which allow for matter with
cosmological symmetry on the brane (Friedmann branes) \cite{BDL}
(in this case the bulk is Schwarzschild-Anti de Sitter space-time
\cite{MSM}), and non-empty bulks have also been considered
including models allowing a \textit{dilatonic} type scalar field
in the bulk \cite{MW}. A geometric formulation of the class of
Randall-Sundrum-type brane-world models was given in~\cite{sms}.
The dynamical equations on the 3-brane differ from the equations in general
relativity (GR) by terms that carry the effects of
imbedding and of the free gravitational field in the 5-dimensional
bulk (transmitted via the projection ${\cal E}_{\mu\nu}$ of the
bulk Weyl tensor) \cite{Maartens}. In general, in the 
4-dimensional picture the conservation equations do not
determine all of the independent components of ${\cal E}_{\mu\nu}$
on the brane (and a complete higher-dimensional analysis,
including the dynamics in the bulk,is necessary).

The asymptotic dynamical evolution of spatially homogeneous
brane-world cosmological models close to the initial singularity,
where the energy density of the matter is larger than the brane
tension and  the behaviour deviates significantly from the
classical general-relativistic  case, was studied in
\cite{chaos,COLEY}. It was found that for perfect fluid models 
with a linear barotropic
$\gamma$-law equation of state an isotropic singularity
\cite{GW85} is a past-attractor in all  orthogonal Bianchi models
and is a local past-attractor in a class of  
inhomogeneous brane-world models for all $\gam>1$. The early
investigations of the initial singularity used only isotropic
fluids as a source of matter \cite{isot}. However, the study of
the behaviour of spatially homogeneous brane-worlds close to the
initial singularity in the presence of both local and nonlocal
stresses indicates that for physically relevent values of the
equation of state parameter there exist two local past attractors
for these brane-worlds, one isotropic and one anisotropic
(although the anisotropic models are likely unphysical and can be
ruled out). In particular, Barrow and Hervik \cite{BH} studied a
class of Bianchi type I brane-world models with a pure magnetic
field and a perfect fluid and found that when $\gamma \ge
\frac{4}{3}$ the equilibrium point ${\cal F}_b$ is again a local
source (past-attractor), but that there exists a second
equilibrium point, which corresponds to a new brane-world solution
with a non-trivial magnetic field, which is also a local source
(and is the only local source when $\gamma < \frac{4}{3}$). This
was generalized by \cite{hervik}, in which it was shown that for a
class of spatially homogeneous brane-worlds with anisotropic
stresses, both local and nonlocal, the brane-worlds could have
either an isotropic singularity or an anisotropic singularity for
$\gam>4/3$.

The {\em governing field equations} induced on the brane, using
the Gauss-Codazzi equations, matching conditions and $Z_2$
symmetry are given in \cite{sms,Maartens}. The energy-momentum
tensor of the matter fields can be decomposed covariantly with
respect to a chosen 4-velocity (timelike vector) $u^\mu$, in terms
of the energy density $\rho$ and isotropic pressure $p$, and
$h_{\mu\nu}=g_{\mu\nu}+u_\mu u_\nu$ projects orthogonal to
$u^\mu$. We are particularly interested in the physical case of a
perfect fluid with $\rho=(\gamma-1)p$, especially with $\gamma \ge
4/3$, where $\gamma = 4/3$ corresponds to radiation and $\gamma
=2$ corresponds to massless scalar fields close to the initial
singularity. The dynamical equations on the 3-brane differ from
the GR equations~\cite{sms,Maartens} in that there are nonlocal
effects from the free gravitational field in the bulk, transmitted
via the projection ${\cal E}_{\mu\nu}$ of the bulk Weyl tensor,
and  the local quadratic energy-momentum corrections, which are
significant at very high energies and particularly close to the
initial singularity. Nonlocal effects from the bulk can be
irreducibly decomposed in terms of an effective nonlocal
energy density, an effective nonlocal anisotropic
stress and an effective
nonlocal energy flux on the brane, ${\cal U}$, ${\cal P}_{\mu\nu}$
and ${\cal Q}_\mu$,
respectively ~\cite{Maartens}.

All of the bulk corrections may be consolidated into an effective
total energy density, pressure, anisotropic stress and energy
flux, so that the modified Einstein equations take the standard
Einstein form with a redefined energy-momentum tensor:
\begin{equation}
G_{\mu\nu}= \kappa^2 T_{\mu\nu}+\widetilde{\kappa}^4S_{\mu\nu} -
{\cal E}_{\mu\nu} \equiv T^{\rm tot}_{\mu\nu} = \rho^{\text{tot}}
u_\mu u_\nu+p^{\text{tot}}h_{\mu\nu}+\pi^{\text{tot}}_{\mu\nu}
+2q^{\text{tot}}_{(\mu}u_{\nu)}\,, \ee where $\kappa^2=8\pi/M_{\rm
p}^2\,,$ ~~$\lambda=6{\kappa^2}/{\widetilde\kappa^4}$, and
\begin{eqnarray}
\rho^{\rm tot} &=& \kappa^2\,\rho+{6\kappa^2 \over
\lambda}\left[{1\over 24}\left(2\rho^2 -3
\pi_{\mu\nu}\pi^{\mu\nu}\right) + {1\over
\kappa^4}{\cal U}\right]\label{a}\\
p^{\rm tot} &=&\kappa^2\, p+ {6\kappa^2\over \lambda}\left[{1 \over
24}\left(2\rho^2+4\rho p+
\pi_{\mu\nu}\pi^{\mu\nu}-4q_\mu q^\mu\right) +{1 \over
3}{1\over\kappa^4}{\cal U}\right] \label{b}\\
 \pi^{\rm tot}_{\mu\nu} &=&
\kappa^2\,\pi_{\mu\nu}+ {6\,\kappa^2\over\lambda}\left[{1\over
12}\left(-(\rho+3p)\pi_{\mu\nu}-3
\pi_{\alpha\langle\mu}\pi_{\nu\rangle}{}^\alpha+3q_{\langle\mu}q_
{\nu\rangle}\right) +{1\over \kappa^4}{\cal P}_{\mu\nu}\right]\label{c}\\
 q^{\rm tot}_\mu &=&\kappa^2\,q_\mu+ {6\,\kappa^2\over\lambda}\left[{1
\over 12}\left(2\rho
q_\mu-3\pi_{\mu\nu}q^\nu\right)+ {1\over \,\kappa^4}{\cal Q}_\mu
\right]\label{d}
\end{eqnarray}

As a consequence of the form of the bulk energy-momentum tensor
and of $Z_2$ symmetry, it follows \cite{sms} that the brane
energy-momentum tensor separately satisfies the conservation
equations, i.e., $\nabla^\nu T_{\mu\nu}=0$. Consequently, the
Bianchi identities on the brane imply that the projected Weyl
tensor obeys the non-local constraint $\nabla^\mu{\cal
E}_{\mu\nu}=\widetilde{\kappa}^4\nabla^\mu S_{\mu\nu}$.
The evolution of the anisotropic stress part is {\em not}
determined on the brane. The correction terms must be consistently
derived from the higher-dimensional equations. The fact that since
${\cal P}_{\mu\nu}$ corresponds to gravitational waves in
higher-dimensions it is expected that the dynamics will not be
affected significantly at early times close to the singularity
\cite{waves}. Henceforth we shall  assume that ${\cal
P}_{\mu\nu}=0$. A dynamical argument to support this assumption is
given in \cite{CHL}. No further assumptions on the models are
made.

From numerical and dynamical considerations it was concluded that \cite{CHL} the area
expansion rate increases without bound (and hence the Hubble rate
$\rightarrow \infty$) as logarithmic time $t \rightarrow -\infty$,
and hence there always exists an {\em initial singularity}. In
addition, the normalized frame variable \cite{elst} vanishes  as
$t \rightarrow -\infty$. This allows us to calculate the
exponential decay rates close to the initial singularity.  The
constrained evolution system of equations for general
inhomogeneous ($G_0$) brane world models were given in
\cite{lim0306118} in terms of the variables $\Omega_b,{\bf X}$ (in
the separable volume gauge using Hubble-normalized equations),
where {\bf X} represents the independent variables corresponding
to the shear and curvature and additional matter terms,  and
$\Omega_b \equiv \mu_b/H^2$, where essentially ${\mu}_b \sim
{{\rho}_b}^2$. The exponential decay rates in the case $\gam>4/3$
were calculated by Lim  (see the Appendix in \cite{CHL}). It was
found that as $t \rightarrow -\infty$ the variables
$(\Omega_b-1),{\bf X}$ have decay rates that depend linearly on $\{\e^{(3\gam-1)t},
\e^{3\gam t}, \e^{3(\gam-1)t} \e^{2(3\gam-4)t}, \e^{2t},
\e^{(3\gam-4)t}, \e^{2(3\gam-2)t}\}$. There is also evidence of
isotropization (albeit slowly) in the degenerate case $\gam=4/3$
(see below). This supports the possibility that in general brane
world cosmologies have an isotropic singularity for $\gam \ge 4/3$.

The evolution of models with an isotropic cosmological 
initial singularity is approximated by the flat
model corresponding to the `equilibrium state' ${\cal F}_b$,
characterized by $\Omega_b = 1$, $\bf{X}= \bf{0}$, which
corresponds to a self-similar, spatially homogeneous and isotropic
non-general-relativistic brane-world model \cite{BDL}. This is
consistent with previous work in which it was shown that for all
physically relevant values of $\gamma$, ${\cal F}_b$ is a
past-attractor in non-tilting  spatially homogeneous brane-world
models \cite{chaos}, and ${\cal F}_b$ is a past-attractor in the family
of spatially inhomogeneous `non-tilting' $G_2$ cosmological models
\cite{chaos}. The results are also consistent with previous
results in the spatially homogeneous orthogonally transitive {\em
tilting} Bianchi type VI$_0$  and VII$_0$ models \cite {Hewitt},
and Bianchi type VI$_0$ models with magnetic field \cite{lkw}
(note, especially, the bifurcation at $4/3$).
It follows immediately that the total energy density $\ti{\rho}
\rightarrow \infty$ as $ t \rightarrow -\infty$ \cite{chaos}, so
that ${\mu}_b \sim {{\rho}_b}^2$ dominates as $ t \rightarrow
-\infty$ and that all of the other contributions to the brane
energy density are negligible dynamically as the singularity is
approached. Hence
close to the singularity the matter contribution is given by
\begin{equation}
{\rho}^{\rm tot} = {1\over 2\lambda}{\rho}^2 \equiv {\mu}_b ;
~~~{p}^{\rm tot} = {1\over 2\lambda}({\rho}^2+2{\rho} {p}) =
(2\gamma -1){\rho}^{\rm tot},
\end{equation}
so that the effective equation of state at high
densities is $(2\gamma -1)$, which is greater than unity in cases
of physical interest.

More detailed information has recently been obtained from a study of
the dynamics of a class of {\em spatially
inhomogeneous $G_{2}$} cosmological models with one spatial degree
of freedom in the brane-world scenario \cite{CHL}. The $G_{2}$
cosmological models admit a 2-parameter Abelian isometry group
acting orthogonally transitively on spacelike 2-surfaces. The
formalism of \cite{elst} was utilized, in which area expansion
normalized scale-invariant dependent variables (the area expansion
rate is effectively the Hubble rate close to an
initial singularity), the timelike area gauge and an effective
logarithmic proper time $t$ were employed, and the initial
singularity occurs for $t\rightarrow -\infty$. The resulting
governing system of evolution equations of the spatially
inhomogeneous $G_{2}$ brane cosmological models is then written
as a system of autonomous first-order partial differential
equations in two independent variables. In terms of
scale-invariant dependent variables, the equations consist of an
{\em Evolution system\/}: $\partial_t\{
{\bf{X}},\Omega_b\}={\bf{F}}({\bf{X}},\Omega_b,
\partial_x{\bf{X}},\partial_x{\Omega_b};\gamma )$, and
{\em Constraint equations and defining equations}: ${\bf G}
({\bf{X}},\Omega_b,
\partial_x{\bf{X}},\partial_x{\Omega_b};\gamma )=0$ \cite{CHL}.
The asymptotic evolution of the class of orthogonally transitive
$G_{2}$ cosmologies near the cosmological initial singularity can
then be discussed. Since the normalized frame variable 
was found to vanish asymptotically,
the singularity is characterized by the fact that spatial
derivatives are dynamically negligible.

The local dynamical behaviour of this class of spatially
inhomogeneous models close to the singularity was then studied
{\em numerically} \cite{CHL}. It was found that the area expansion
rate increases without bound (and hence the Hubble rate
$H \rightarrow \infty$) as $t \rightarrow -\infty$,
so that there always exists an initial singularity. For $\gam
>4/3$, the numerics indicate $\{\bf{X}\} \rightarrow 0$ as
$t\rightarrow -\infty$ (and $\Om_b \rightarrow 1$) for {\em all}
initial conditions. In the case of radiation ($\gam=4/3$), the
models were still found to isotropize as $t\rightarrow -\infty$,
albeit slowly. For $\gam<4/3$, $\{ \bf{X} \}$ tend to constant but
not necessarily zero values as $t\rightarrow -\infty$. In fact, the
numerical results support the fact that {\em  all} cosmological
models have an isotropic singularity for {\em  $\gamma > 4/3$} (i.e.,
the singularity is isotropic for all initial
conditions, indicating that ${\cal F}_b$) is a
global past-attractor).

From the numerical analysis we conclude that there is an initial
isotropic singularity  in all of these $G_2$ spatially
inhomogeneous brane cosmologies for a range of parameter values
which include the physically important cases of radiation and a
scalar field source. The numerical results are supported by a
qualitative dynamical analysis and a detailed calculation of the
past asymptotic decay rates. Although the analysis is local in
nature, the numerics indicates that the singularity is isotropic
for all initial conditions for the range of parameter values of
physical import \cite{CHL}.

These results have been further supported by a detailed
analysis of {\em linear perturbations} of the isotropic brane
model ${\cal F}_b$ using the covariant and gauge invariant
approach \cite{DGBC}. In particular, a detailed analysis of
generic linear inhomogeneous and anisotropic perturbations
\cite{EB} of the past attractor ${\cal F}_b$ was carried out by
deriving a full set of linear 1+3 covariant {\it propagation} and
{\it constraint} equations for this background, split into scalar, 
vector and tensor parts, which govern the complete
perturbation dynamics of the physical quantities that describe the
kinematics of the fluid flow and the dynamics of the gravitational
field. The analysis was restricted to large scales, at a time when
physical perturbation scales are much larger than the Hubble
radius, $\lambda\gg H^{-1}$, which is of relevance for the
discussion of non-inflationary perfect fluid models. In fact,
since any wavelength $\lambda<H^{-1}$ at a given time becomes much
larger than $H^{-1}$ at earlier times, 
this perturbation analysis is completely general. Solutions to
the set of perturbation equations were presented in Table I in
\cite{DGBC}, where it was concluded that ${\cal F}_b$ is stable in
the past to generic inhomogeneous and anisotropic perturbations
for physically relevant values of $\gamma$. In addition, it follows
immediately that the expansion normalised shear vanishes as as the
initial singularity is approached, and isotropization occurs.

We have argued that in spatially inhomogeneous
brane-world cosmological models the initial singularity is {\em
isotropic\/} ~\cite{GW85}. Therefore, unlike the situation in GR,
it is plausible that typically the initial singularity is
isotropic in brane world cosmology. Such a `quiescent'
cosmology~\cite{Barrow}, in which the universe began in a highly
regular state but subsequently evolved towards irregularity, might
offer an explanation of why our Universe might have began its
evolution in such a smooth manner and may provide a realisation of
Penrose's ideas on gravitational entropy  and the second law of
thermodynamics in cosmology~\cite{Penrose79}. More importantly, it
is therefore possible that a quiescent cosmological period
occuring in brane cosmology provides a physical scenario in which
the universe starts off smooth and that naturally gives rise to
the conditions for inflation to subsequently take place.
Cosmological observations indicate that we live in a Universe
which is remarkably uniform on very large scales. However, the
spatial homogeneity and isotropy of the Universe is difficult to
explain within the standard GR framework since, in the presence of
matter, the class of solutions to the Einstein equations which
evolve towards a RW universe is essentially a set of measure zero.
In the inflationary scenario, we live in
 an isotropic region of a potentially highly
irregular universe as the result of an expansion phase in the early universe
thereby solving many of the problems of cosmology. Thus this
scenario can successfully generate a homogeneous and
isotropic RW-like universe from initial conditions which, in the
absence of inflation, would have resulted in a universe far
removed from the one we live in today. However, still only a restricted set
of initial data will lead to smooth enough conditions for the
onset of inflation.

Let us discuss this in a little more detail. Although inflation
gives a natural solution of the horizon problem of the big-bang
universe, inflation requires homogeneous initial conditions over
the super-horizon scale, i.e., it itself requires certain
improbable initial conditions. When inflation begins to act, the
universe must already be smooth on a scale of at least $10^5$
times the Planck scale. Therefore, we cannot say that it is a
solution of the horizon problem, though it reduces the problem by
many orders of magnitude.  Many people have investigated how
initial inhomogeneity affects the onset of inflation, and it was
found that including spatial inhomogeneities accentuates the
difference between models like new inflation and those like
chaotic inflation.  Goldwirth and Piran \cite{GP}, who solved the
full Einstein equations for a spherically symmetric  spacetime,
found that {\em small-field} inflation models of the type of {\em
new inflation} is so sensitive to initial inhomogeneity that it
requires homogeneity over a region of several horizon sizes. {\em
Large-field} inflation models such as {\em chaotic inflation} is
not so affected by initial inhomogeneity but requires a
sufficiently high average value of the scalar field over a region
of several horizon sizes \cite{Brandenberger}. Therefore,
inhomogeneities further reduce the measure of initial conditions
yielding new inflation, whereas the inhomogeneities have
sufficient time to redshift in chaotic inflation, letting the zero
mode of the field eventually drive successful inflation. In
conclusion, although inflation is a possible causal mechanism for
homogenization and isotropization, there is a fundemental problem
in that the initial conditions must be sufficiently smooth in
order for inflation to subsequently take place \cite{COLEY}. We
have found that  an isotropic singularity in brane world cosmology
might provide for  the
 necessary sufficiently smooth initial
conditions to remedy this problem.

{\em Acknowledgements}

This work was supported by the Natural Sciences and Engineering
Research Council of Canada.

\end{document}